 \documentclass[twocolumn]{aa}
 \usepackage{graphicx}
\usepackage{url}
 \usepackage{tabularx}

 \usepackage{txfonts}
\begin{document}

    \title{Multi-wavelength study of the gravitational lens system RXS~J113155.4-123155}

    \subtitle {I. Multi-epoch optical and near infrared imaging
    \thanks{Based on observations collected at the European Southern
    Observatory, La Silla and Paranal, Chile (ESO Program
    71.A-0407(A, E) and DDT Program 272.A-0535(A)); on observations obtained with the NASA/ESA Hubble Space Telescope
    (Space Telescope Science Institute) which is operated by the
    Association of Universities for Research in Astronomy, Inc., under
    NASA contract NAS 5-26555 (program \#9744); and on observations obtained at the Canada-France-Hawaii Telescope (CFHT) which is operated by the National Research Council of Canada, the Institut National des Sciences de l'Univers of the Centre National de la Recherche Scientifique of France,
and the University of Hawaii.}}

    \author{D. Sluse\inst{1,2,3}, J.-F. Claeskens\inst{1}, B. Altieri
    \inst{4}, R.A. Cabanac \inst{5}, O. Garcet \inst{1},
    D. Hutsem\'ekers\inst{1}\thanks{Chercheur qualifi\'e du
    F.N.R.S. (Belgique)}, C. Jean \inst{1}, A. Smette
    \inst{1,2}$^{\star\star}$ \and J. Surdej\inst{1}\thanks{Directeur de
    recherches honoraire du F.N.R.S. (Belgique)}}

    \offprints{dominique.sluse@epfl.ch}

    \institute{Institut d'Astrophysique et de G\'eophysique, Universit\'e de Li\`ege,
    All\'ee du 6 Ao\^ut 17, B5C, B-4000 Sart Tilman, Belgium \and
    European Southern Observatory, Alonso de Cordova 3107, Santiago
    19, Chile \and Laboratoire d'Astrophysique, Ecole Polytechnique F\'ed\'erale de Lausanne
    (EPFL) Observatoire, 1290 Sauverny, Switzerland \and European Space
    Astronomy Centre, ESA, P.O. Box 50727, 28080 Madrid, Spain \and
    Canada-France-Hawaii Telescope, 65-1238 Mamalahoa Highway,
    Kamuela, HI 96743, USA }

    \date{Received: ;  accepted: }

 
  \abstract
   {}
   {RXS J113155.4-123155 (z=0.66) is a quadruply imaged lensed
     quasar with a resolved Einstein Ring. The goal of this paper is
     to provide a full characterization of this system, and more
     particularly accurate astrometry and photometry. These
     observational constraints constitute a mandatory ingredient for
     the precise determination of the lens mass profile, the
     derivation of the Hubble constant $H_0$ from time delay
     measurements and investigations on the presence of massive
     substructures in the lensing galaxy.}
   {Visible and near-infrared imaging observations of RXS
     J113155.4-123155 were carried out at various epochs using several
     ground based telescopes and the HST. The frames have been
     deconvolved using the MCS algorithm. A
     Singular Isothermal Ellipsoid (SIE) + external shear has been
     used to model the lensing galaxy potential. }
   { MCS deconvolution enables us to separate the flux of the QSO
     (point-like images) from that of its host galaxy and to
     accurately track the flux variations of the point-like images in
     various filters. The deconvolved frames unveil several multiply
     imaged structures in the Einstein ring and an unidentified object
     in the vicinity of the lensing galaxy.  We discuss the
     lightcurves and the chromatic flux ratio variations and deduce
     that both intrinsic variability and microlensing took place
     during a span longer than one year. 
     
     We demonstrate that microlensing may easily account for the so
     called anomalous flux ratios presented in the discovery paper.
     However, the observed flux ratios are still poorly reproduced
     when modeling the lens potential with a SIE+shear. We argue that
     this disagreement can hardly be explained by milli-lensing caused
     by substructures in the lensing galaxy. A solution proposed
     in Paper II consists in a more complex lens
     model including an octupole term to the lens gravitational
     potential.  }
   {}
   
    \titlerunning{Multi-wavelength and multi-epoch imaging of RXS
    J113155.4-123155.}  \authorrunning{Sluse D. et al.}

 
    \maketitle

 %

 \section{Introduction}
\label{sec:intro}

RXS J113155.4-123155 (hereafter J1131) is one of the nearest confirmed
multiply imaged AGN. The source at $z_s = 0.658$ is lensed by an
elliptical galaxy at $z_l = 0.295$ (Sluse et al. \cite{SLU03};
hereafter {\it{Paper I}}). This system is a long axis quad with an
image configuration very similar to B1422+231 (Patnaik et
al. \cite{PAT92}): three merging images (B-A-C; typical of a source
lying close to a cusp caustic) face the faint saddle-point image D
lying close to the lensing galaxy (G). For such a system (i.e. cusp
configuration system), the magnification behaviour is well understood
- the flux of the middle image should be equal to the total fluxes of
the two outer images (Schneider \& Weiss \cite{SCH92a}). However, this
generic prediction was strongly violated at the epoch of the
discovery, suggesting the likely presence of small-scale structures in the
lensing galaxy (Keeton, Gaudi \& Petters ~\cite{KEE03b},
~\cite{KEE05}; hereafter KGP03, KGP05). Thanks to the data set presented here, we re-examine
the values of the flux-ratios with regard to their temporal and chromatic
variations. Additionally, because this quad is quite bright and shows
lensed images with a wide angular separation ($2 \theta_E
\sim$ 3.6\arcsec), it is potentially a target of interest for time
delay measurements. Also, the bright Einstein ring detected from the
optical to the near-infrared offers unique constraints on the lens
modeling (Kochanek et al.~\cite{KOC01}) that may break the degeneracy
between the intrinsic lens galaxy shape and the external shear (Keeton et al.
\cite{KEE00}). However, the bright Einstein ring introduces an
additional complexity to the photometry measurement because it
is superimposed over the lensed point-like images. This motivates our extensive
discussion of the sources of systematic photometric errors affecting
the individual lensed QSO images and of the photometric accuracy that
can be reached for J1131.

The observations presented here constitute the first part of a
follow-up study of this system that also includes detailed lens models
and source reconstruction from HST imaging, X-ray imaging as well as
optical and NIR spectroscopy. Sects.~\ref{sec:Obs}
\&~\ref{sec:astrophot} describe the data reduction and
analysis. Section~\ref{sec:results} includes a description of the ring
morphology, a simple modeling of the lensing galaxy and a discussion
of the flux (time and chromatic) variability of the lensed
images. Finally, Sect.~\ref{sec:conclusions} summarizes the main
results and presents our conclusions. 

We adopt throughout this paper $H_0=$ 65~km s$^{-1}$ Mpc$^{-1}$,
$\Omega_0=$ 0.3 and $\Lambda_0=$ 0.7; magnitudes are computed in
the Vega system.


 \section{Observations and reductions} 
 \label{sec:Obs}

In this section, we present NIR and optical imaging obtained for
J1131. Near Infrared observations include images recorded with ESO
telescopes, with CFHT and HST. The images recorded at visible
wavelengths have been obtained with the FORS1 and FORS2 instruments at
the Cassegrain focus of Antu (UT1) and Yepun (UT4) in
Paranal. Table~{\ref{tab:log}} summarizes the technical data of the
various instrumentations used and lists the observational characteristics
of each data set.

\begin{table*}[]
\caption{Log of the observations. Epoch = epoch number as used in the
  text. CONAD = electron to ADU conversion 
  factor.  NExp = number of dithered
  frames. Exp = total
  integration time of a single frame. For the NIR data, Exp = NDIT$\times$DIT, where NDIT is the number of
  sub-frames with an integration time of DIT seconds. Only the total
  exposure time per frame is reported for the NIC2 images. FWHM = mean
  seeing measured on frames. Note that the SOFI pixel size has been re-measured
to be 0.289\arcsec, superseding the published ESO value of
0.292\arcsec.}
\begin{center}
\begin{tabular}{cccccccccc}
\hline 
 Date & Epoch & Instrument & RON & CONAD & Pix. size &Filter & NExp & Exp & FWHM\\
 (dd-mm-yyyy) &  & (Telescope) & (e$^-$/pixel)& 
 (e$^-$/ADU) & (\arcsec) & & & (s) & (\arcsec)\\

\hline 
21-11-2002 & 1 & SOFI (NTT) & 11 & 5.3 & 0.289 & $J$ & 24 & 2$\times$30 & 0.7-0.9\\
12-04-2003 & 2 & ISAAC (UT1) & 11 & 4.6 & 0.1484& $Ks$& 48 & 6$\times$10 & 0.3-0.55\\
21-04-2003 & 3 & FORS2 (UT4) & 4.2 & 1.25 & 0.1263&$R$ special& 4 & 80 & 0.65-0.85 \\
02-05-2003 & 4 & FORS2 (UT4) & 4.2 & 1.25 & 0.1263& $R$ special& 6 & 40 & 0.8 \\
26-05-2003 & 5 & FORS2 (UT4) & 4.2 & 1.25 & 0.1263& $B$ Bessel& 8 & 497 & 0.65-0.9\\
26-05-2003 & 5 & FORS2 (UT4) & 4.2 & 1.25 & 0.1263& $B$ Bessel& 7 & 125 & 0.75-1.0\\
17-06-2003 & 6 & FORS2 (UT4) & 4.2 & 1.25 & 0.1263& $R$ special& 6 & 40 & 0.45-0.50 \\
17-06-2003 & 6 & FORS2 (UT4) & 4.2 & 1.25 & 0.1263& $V$ Bessel & 20 & 59 & 0.5-0.65 \\
17-11-2003 & 7 & NIC2  (HST) & 26  & 5.4  & 0.075 & $F160W$ & 5 & 640 & 0.12 \\
17-11-2003 & 7 & NIC2  (HST) & 26  & 5.4  & 0.075 & $F160W$ & 3 & 704 & 0.12 \\
18-11-2003 & 7 & FORS2 (UT4) & 4.2 & 1.25 & 0.1263&$B$ Bessel & 3 & 240 & 1.0 \\ 
18-11-2003 & 7 & FORS2 (UT4) & 4.2 & 1.25 & 0.1263&$V$ Bessel & 4 & 120 & 0.8-1.0 \\
18-11-2003 & 7 & FORS2 (UT4) & 4.2 & 1.25 & 0.1263&$R$ special & 4 & 80 & 0.85-0.95 \\
09-02-2004 & 8 & CFHT-IR (CFHT) & 15 & 2.35 & 0.211 & $J\#5133$ & 4 & 6$\times$120 & 0.6-0.8\\
09-02-2004 & 8 & CFHT-IR (CFHT) & 15 & 2.35 & 0.211 & $H\#5209$ & 4 & 6$\times$60 & 0.6-0.8\\
09-02-2004 & 8 & CFHT-IR (CFHT) & 15 & 2.35 & 0.211 & $K'\#5337$ & 8 & 3$\times$60 & 0.55-0.9\\
12-04-2004 & 9 & FORS1 (UT1) & 5.2 & 1.61 &0.100 & $B$ Bessel & 3 & 60 & 0.7-0.8\\
12-04-2004 & 9 & FORS1 (UT1) & 5.2 & 1.61 &0.100 &$V$ Bessel & 4 & 30 & 0.65-0.85\\
12-04-2004 & 9 & FORS1 (UT1) & 5.2 & 1.61 &0.100 &$R$ Bessel & 4 & 20 & 0.7-0.85\\
12-04-2004 & 9 & FORS1 (UT1) & 5.2 & 1.61 &0.100 &$I$ Bessel & 4 & 20 & 0.6-0.7 \\

\hline
\end{tabular} 
\label{tab:log}
\end{center} 
\end{table*}


\subsection{Ground-based NIR observations}
\label{subsec:ground}

First, $J$ band images of J1131 have been obtained with SOFI. The conditions were
photometric and 6 standard stars have been observed during the night
enabling us to derive the magnitude of an object producing 1 count/s
of 23.15 $\pm$ 0.01 (hereafter zero-point). 

Secondly, images of J1131 have been obtained in the $Ks$-band with the
Short Wavelength Imaging camera of the ISAAC instrument. Thanks to the
short DIT (10s) used, the combined flux from the sky and QSO images in
any pixel never exceeded 10000 ADU, a level below the non-linearity
regime of the ISAAC detector. Conditions were photometric and a
zero-point of 24.17$\pm$0.07 has been deduced from the standard
stars observed during the night.

Finally, a set of NIR images of J1131 has been obtained at a {\it
single} epoch with CFHT-IR through the broad band $J\#5133$,
$H\#5209$ and $K'\#5337$ filters.

For the whole set of data, standard NIR reduction procedures were
applied to subtract the dark and to flat field the images using a
normalized flat field. Sky subtraction and co-addition of the reduced
dithered frames were performed using the {\texttt{xdimsum}}
IRAF{\footnote{IRAF is distributed by the National Optical Astronomy
    Observatories, which are operated by the Association of
    Universities for Research in Astronomy, Inc., under cooperative
    agreement with the National Science Foundation.}} package. The
residual sky pattern has been fitted and subtracted using the
SExtractor software v~2.3.2 (Bertin, {\cite{bertin}}). 


\subsection{Ground-based optical observations}
\label{subsec:HST}

Observations with the FORS2 instrument have been obtained with the
High Resolution (HR) collimator and the CCD detector in the 2$\times$2
binning mode. Conditions were not photometric.

$B$, $V$, $R$, $I$ images of J1131 were obtained in April 2004
under photometric conditions with the FORS1 instrument and a CCD in
the  unbinned mode. Zero-points for the Bessel $B$, $V$, $R$, $I$ filters
were found to be 27.76$\pm$0.02, 28.12$\pm$0.02,
28.06$\pm$0.04 and 27.25$\pm$0.01 respectively.

Standard reduction procedures were applied to the data including bias
subtraction and flat fielding with a normalized flat. The photometric
measurements of the standard stars have been performed with the
{\texttt{phot}} IRAF task. The sky has been subtracted using the SExtractor
software v~2.3.2 (Bertin, {\cite{bertin}}).


\subsection{HST F160W observations}
\label{subsec:HST}

HST F160W observations of J1131 have been obtained on November 17, 2003 with
the {\it Hubble Space Telescope} (HST-GO-9744; PI: C.S. Kochanek) as part of the CfA-Arizona Space
Telescope Lens Survey (CASTLES). Out of the eight dithered frames of
J1131 obtained with the NIC2, two frames have been excluded from our
analysis because of an error flag associated with pixels
located in the brightest point-like images. We reduced the images
using the IRAF {\texttt{calnica}} and {\texttt{calnicb}} tasks as
recommended in the NICMOS data handbook (Dickinson et
al. \cite{DIC02}). The flux calibration (Dickinson et
al. \cite{DIC02}) has been derived from PHOTNU$=$1.4981 10$^{-6}$ Jy
sec DN$^{-1}$ and from the averaged flux density of VEGA $<F_{\nu}> =
1043.5$ Jy (June 2004 values).


\section{Data analysis }
\label{sec:astrophot}

\subsection{The method}
\label{subsec:method}

The Einstein ring of J1131 is the brightest one known to-day. Because
it is superimposed over the lensed QSO images, standard photometric
methods like PSF fitting may bias the point-like flux measurements. We
thus measured relative photometry using the MCS deconvolution method
(Magain et al. \cite{magain}) which proved to be very efficient in
measuring fluxes of point-like components superimposed over a varying
background (Burud et al. \cite{BUR98}). Thanks to this code, images
are deconvolved to an improved but finite resolution using a
    kernel constructed from PSF stars (Courbin et al. \cite{COU98}).
For HST data, the kernel is constructed based on the PSF
generated with the Tiny Tim v 6.1 software (Krist \& Hook
\cite{KRI03}). In the other cases, the kernel is constructed
from relatively isolated bright field stars. Additionally, since the
co-addition of different images often degrades the PSF, we
generally worked on the individual frames obtained at different
dithered positions.

Instead of deconvolving each image separately, we used a modified
version of the code that {\it simultaneously} deconvolves a set of
individual frames taken with the same instrumental setup (Burud et
al. \cite{BUR2000}), leading to better constraints on the astrometry of the quasar images
(an output of the deconvolution process), on the shape of the Einstein
ring and of the lensing galaxy, using the S/N of the whole data
set. For each data set, deconvolved frames are reconstructed to a
common improved resolution. Technically, the relative positions of the
point-like components are fitted on the reference frame while their
flux and the frame offset with respect to the reference vary
independently for each frame. For the data sets with the largest pixel
size (i.e. CFHT-IR and SOFI), MCS could not fit accurately both the offset with
respect to the reference frame and the point-like image positions. In that case,
the latter have been fixed to the ones derived from the HST
observations. 

In the following, we have simultaneously deconvolved all the frames:
i) for each epoch (in a given filter) separately; ii) for all the epochs available (in a
filter). Single epoch deconvolution is best suited to
retrieve absolute photometry (Sect.~\ref{subsec:absphot}) since it is
not affected by the flux scaling between several epochs. It
is also the only method applicable to the $I$ band and NIR filters
(i.e. one epoch data). Multi-epoch deconvolution is more appropriate
to retrieve accurate flux ratios (Sect.~\ref{subsec:relphot}) and to
measure the temporal flux variations (Sects.~\ref{subsec:relphot}
\&~\ref{subsec:BVR}). Typical photometric random errors obtained with MCS
reach the photon noise uncertainty. 

Although the MCS deconvolution enables one to retrieve accurate
photometry, the results are affected by systematic errors (Burud et
al. \cite{BUR98}) mainly associated with the regularization term
$\lambda$. The latter works like a local smoothing term by reducing
the high frequencies of the background. Thus, depending on the choice{\footnote{A high value of $\lambda$ means a small
    regularization and a low value means a strong smoothing of the
    deconvolved image.}} of $\lambda$, a variable fraction of the background flux
is included in the point-like flux measurements. We selected realistic
values of $\lambda$ based on the flatness of the {\it residual map}
(i.e. the difference between the real image and the deconvolved image
re-convolved with the kernel, and divided by the standard
deviation in each pixel of the real image). More specifically,
    the higher level of regularization is picked when we observe
    more than 3 $\sigma$ standard deviation in the residual map
    (regularization is swamping -true- structures existing in the
    original image) and the lower level is chosen when important
    Gibbs oscillations appear in the deconvolved frame. In the
    following, we run MCS with three acceptable levels of
    regularization: a lower and upper bound -resp. $\lambda_{\rm low}$
    and $\lambda_{\rm high}$- and an intermediate value $\lambda_{\rm
      best}$ considered as the best value.

Another source of systematic photometric errors may come from an
imperfect PSF model. However, this appears as deviations from a null
residual map and the associated photometric bias is corrected by means
of the comparison of deconvolution and aperture photometry of field
stars (see e.g. Sect.~\ref{subsec:absphot}). Note that in such
    cases, since the residual map cannot be used to pick up $\lambda_{\rm
      low}$ and $\lambda_{\rm high}$, we select them based on the
    visual inspection of the deconvolved frame: for a too small value of
    $\lambda$, the background does not look like a ring+lens galaxy,
    but just like a very diffuse background while we determine
    $\lambda_{\rm high}$ via the appearance of important Gibbs
    oscillations.

By default, the error bars reported in this paper are 1$\sigma$
standard errors on the mean.


\subsection{Astrometry}
\label{subsec:relastr}

\begin{table}[tb]
\caption{Relative positions of the different lensed components (BCD),
  the lensing galaxy (G) and the companion object (X) with respect to
  A. These are deduced from a gaussian fitting (except X with MCS) on the HST NIC2 image ($F160W$)
  and from the MCS deconvolution of the FORS2 and ISAAC frames. The last column lists the
  1$\sigma$ random error in both right ascension and
  declination. For the ground-based data, the error on the
    relative position is reported. 1$\sigma$ systematic errors related to scaling, orientation and
    distortion uncertainties are estimated to be 0.003'' (see text
    for details).}
\begin{center}
\begin{tabular}{c|rr|rr|cc}
\hline 
ID & \multicolumn{2}{c}{$\Delta \alpha \cos \delta('')$} & \multicolumn{2}{c}{$\Delta \delta ('')$} & \multicolumn{2}{c}{1$\sigma (\arcsec)$} \\
\hline 
& F160W & ground & F160W & ground & F160W & ground \\
\hline
A & 0 & 0 & 0 & 0 & 0.001 & - \\
B & 0.032 & 0.030 & 1.184 & 1.190 & 0.001 & 0.001 \\
C & $-$0.590 & $-$0.589 & $-$1.117 & $-$1.117 & 0.001 & 0.002 \\ 
D & $-$3.115 & $-$3.120 & 0.875 & 0.881 & 0.001 & 0.002 \\
G & $-$2.027 & $-$2.027 & 0.607 & 0.604 & 0.001 & 0.002 \\
X & $-$1.936 & - & 1.100 & - & 0.011 & - \\
\hline
\end{tabular} 
\label{tab:astro}
\end{center} 
\end{table}

The relative astrometry is derived from the HST data using two
methods. First, we performed 2D gaussian fitting of the individual
lensed images and of the lensing galaxy (each component isolated in a
box of 13$\times$13 pixels) with the {\texttt{i2gaussfit}} IRAF
task. Mean positions and associated errors are calculated from the
measurements on the 6 individual frames.  
 
In the second method, the relative astrometry of images B, C, D and of
the lens galactic nucleus G relative to A is derived {\it simultaneously}
using MCS. Unfortunately, this method suffers from the poor PSF
sampling and is very sensitive to the prior knowledge of the relative
offset between the frames. Hence, we have used the gaussian fitting
results to fix the mean relative offset and measured the mean position
of object {\it X} (see Sect.~\ref{subsec:ring}) with MCS. Note that
MCS positions of A, B, C, D and G are in statistical agreement with
the gaussian fitting results.

The measured positions have been converted to the RA/DEC system
using a plate scale{\footnote
  {\url{http://www.stsci.edu/hst/nicmos/performance/platescale}.}} of
0.07588\arcsec/pixel in x and of 0.07537 \arcsec/pixel in y and a
rotation angle of 59.202$\degr$ (E of N).

Relative astrometry is also derived from the best ground-based
observations (i.e. FORS2 and ISAAC data). Since the spatial sampling
of the ground-based PSFs fully satisfies the sampling theorem, MCS
leads to an accurate relative astrometry. The relative positions of
the lensed quasar images and of the lensing galaxy found in each band
(namely $B$, $V$, $R$ and $Ks$) do agree between each other,
  within typically less than 0.01\arcsec. The standard deviation of
the positions measured in each filter is used to derive the 1$\sigma$
standard error on the mean relative positions. However, because of the
small separation of the triplet of images A, B and C, the measured
positions of B and C are likely correlated with that of A, resulting
in an underestimate of the error. Consequently, we have arbitrarily
and quadratically added a 0.001'' error to the 1$\sigma$ standard errors
of B and C.

The systematic error on the position is estimated to be
0.003$\arcsec$. This error is typical of the systematic error that may
affect the relative astrometry of small separation systems on NIC2
data (Impey et al. \cite{IMP98}). For the ground-based data, the
systematic error is inferred from the recalibration of the pixel size
of the FORS frames with the {\texttt {MIDAS}} implementation of
pos1 (Walter \& West \cite{WAL86}) and with {\texttt
  {GAIA}}{\footnote{{\texttt {GAIA}} is a
    derivative of the Skycat catalog and image display tool, developed
    as part of the VLT project at ESO. Skycat and GAIA are free
    softwares under the terms of the GNU
    copyright.}} (v~2.5.3). Table~{\ref{tab:astro}} summarizes the
relative positions we derived for J1131 based on the HST and
ground-based data. Both sets of results do agree within
0.006$\arcsec$.


\subsection{Photometry}

\begin{table*}[!t]
\caption{Photometry in the $B$-$V$-$R$-$I$ Bessel filters (epoch 9),
  in the $J$ band (epoch 1), in the $H$-band ($F160W$ Vega normalized;
  epoch 7) and in the $Ks$ band (epoch 2) for images A, B, C and D of
  J1131 as obtained with MCS. The notation is $m_{\rm best} \pm
  {{\sigma}_{\rm random}}~^{+{\rm sys}}_{-{\rm sys}}$, where $m_{\rm
    best}$ is the mean value obtained for $\lambda_{\rm best}$,
  ${\sigma}_{\rm random}$ is the quadratic sum of the 1$\sigma$ error
  on the mean, the error on the extinction   coefficient and the error
  on the zero-point (this one amounts to 0.03 mag for NICMOS
  photometry) and $\pm {\rm sys}$ is the peak to peak variation
  between extreme values of $\lambda$. Last column gives the FWHM (in
  arcsec) of the PSF in the deconvolved frame. When null error bars
  appear in the table, the error bar is of only a few milli-magnitudes.}

\begin{center}
\begin{tabular}{ccccccccccc}
\hline 

 Filter & epoch & A & B & C & D & FWHM\\
\hline 
\vspace{2pt}
$B$ & 9 & $18.09\pm0.02^{+0.00}_{-0.00}$ &
$18.14\pm0.02^{+0.00}_{-0.00}$ & $19.19\pm0.02^{+0.02}_{-0.01}$ &
$20.64\pm0.02^{+0.03}_{-0.01}$ & 0.30 \\
\vspace{2pt}
$V$ & 9 & $17.88\pm0.02^{+0.02}_{-0.01}$ &
$17.83\pm0.02^{+0.03}_{-0.01}$ & $18.96\pm0.02^{+0.07}_{-0.00}$ &
$20.41\pm0.02^{+0.11}_{-0.06}$ & 0.30 \\
\vspace{2pt}
$R$ & 9 & $17.76\pm0.04^{+0.02}_{-0.02}$ &
$17.66\pm0.04^{+0.04}_{-0.03}$ & $18.77\pm0.04^{+0.10}_{-0.05}$ &
$20.21\pm0.04^{+0.04}_{-0.06}$ & 0.35 \\
\vspace{2pt}
$I$ & 9 & $17.43\pm0.01^{+0.05}_{-0.01}$ &
$17.42\pm0.01^{+0.07}_{-0.02}$ & $18.44\pm0.01^{+0.14}_{-0.02}$ &
$19.72\pm0.01^{+0.03}_{-0.03}$ & 0.30 \\
\vspace{2pt}
$J$ & 1 & $16.83\pm0.03^{+0.05}_{-0.01}$&
$16.49\pm0.03^{+0.03}_{-0.00}$&$17.48\pm0.04^{+0.04}_{-0.02}$ &
$19.00\pm0.04^{+0.02}_{-0.01}$ & 0.29 \\
\vspace{2pt}
$F160W$ & 7 & $15.80\pm0.04^{+0.04}_{-0.17}$ &
$16.08\pm0.03^{+0.03}_{-0.02}$ & $16.95\pm0.04^{+0.21}_{-0.04}$ &
$18.36\pm0.07^{+0.03}_{-0.02}$ & 0.08\\
\vspace{2pt}
$Ks$ & 2 & $14.72\pm0.07^{+0.01}_{-0.00}$ &
$15.11\pm0.07^{+0.00}_{-0.01}$ & $15.72\pm0.07^{+0.02}_{-0.02}$ &
$17.17\pm0.07^{+0.02}_{-0.03}$ & 0.15\\
\hline
\end{tabular} 
\label{tab:absphot}
\end{center} 
\end{table*}

\subsubsection{Absolute photometry}
\label{subsec:absphot}

We derive absolute photometric measurements with MCS in the
Bessel $B$, $V$, $R$, $I$ bands and for the $J$, $F160W$ and $Ks$
filters (Table~{\ref{tab:absphot}}) based on the fluxes obtained with
$\lambda_{\rm best}$.  Sets of individual frames are used except for
the $J$-SOFI ($Ks$) filters where a set of 4 (8) stacked frames of 6$\times$1
minutes is considered. The peak to peak systematic variations
introduced by $\lambda$ are estimated from the results obtained
for $\lambda_{\rm low}$ and $\lambda_{\rm high}$. Fluxes are converted into the Vega
magnitude system using the zero-points reported in
Sect.~{\ref{subsec:ground}} and applying an extinction correction (see 
the ESO webpages for extinction coefficients). Additionally, for
each set of ground-based images, we compared the aperture photometry obtained with the
{\texttt{phot}} IRAF routine and SExtractor to MCS
photometry of at least 4 stars in the field. All the measurements are
consistent, suggesting the absence of systematic errors caused by the
estimate of the PSF. 

For the HST NIC2 data, the Tiny Tim PSF poorly reproduces the
observations. Consequently, we find several acceptable solutions for
each possible value of $\lambda$ ($\lambda_{\rm best}$ but also
$\lambda_{\rm low, high}$) suggesting that the minimization routine in MCS is
trapped in one of the numerous local minima. Hence, the random error
bars of the F160W photometry presented in Table~{\ref{tab:absphot}}
are likely underestimated and the associated peak to peak variations
may also be biased. Although we do not know by which amount, we trust
these results since they are found in good agreement with independent PSF
fitting photometry presented in Claeskens et al. (accepted;
hereafter {\it Paper II}).  


\subsubsection{Relative photometry}
\label{subsec:relphot}

\begin{table*}[tb]
\caption{Magnitude differences (Sect.~\ref{subsec:relphot}). The notation is~$m_{\rm best}
  \pm {{\sigma}_{\rm random}}~^{+{\rm sys}}_{-{\rm sys}}$, where~
  $m_{\rm best}$ is the mean value obtained for $\lambda_{\rm best}$,
  ${\sigma}_{\rm random}$ is the 1$\sigma$ error on the mean and $\pm
  {\rm sys}$ refers to the peak to peak variations between extreme values of
  $\lambda$. {\bf {Col 1}}: filter; {\bf {Col 2}}: epoch for which the 
    magnitude differences are given; {\bf {Col 3}}: R$_{ij}$(A) = $I_i/I_j$
    for image A. $i$ is the reference epoch (bold face)
    for the filter considered while $j$ is the epoch given in
    Col. 2. {\bf {Col. 4-5-6}}: average magnitude differences between image A
    and resp. B, C, D. {\bf {Col. 7-8}}: idem between image B and resp. C,
    D. {\bf {Col. 9}}: FWHM (in arcsec) of the PSF in the deconvolved
    frame. Since the flux
    variations (Col. 3) are not affected by the regularization,
    systematic error bars are not expressed.   When null error bars
    appear in the table, the error bar is only of a few milli-magnitudes.  }
\begin{center}
\begin{tabular}{ccccccccc}
\hline 
 Filter & E & R$_{ij}$(A) & $\Delta m_{\rm AB}$ & $\Delta m_{\rm
   AC}$ & $\Delta m_{\rm AD}$ & $\Delta m_{\rm BC}$ & $\Delta m_{\rm
   BD}$ & FWHM \\
\hline 
\vspace{2pt}
$B$ & {\bf 5} & - & $0.19\pm0.00^{+0.00}_{-0.00}$ &
$-0.92\pm0.00^{+0.00}_{-0.00}$ & $-2.31\pm0.00^{+0.01}_{-0.01}$ &
$-1.11\pm0.00^{+0.00}_{-0.00}$ & $-2.50\pm0.00^{+0.01}_{-0.01}$ & 0.30 \\
\vspace{2pt}
$B$ & 7 & $0.96\pm0.04$ & $-0.01\pm0.00^{+0.00}_{-0.00}$ & $-1.06\pm0.01^{+0.00}_{-0.00}$ & $-2.36\pm0.00^{+0.00}_{-0.01}$ & $-1.05\pm0.01^{+0.00}_{-0.00}$ & $-2.35\pm0.00^{+0.01}_{-0.01}$  & 0.30 \\
\vspace{2pt}
$B$ & 9 & $1.00\pm0.04$ & $-0.05\pm0.00^{+0.00}_{-0.00}$ & $-1.11\pm0.00^{+0.00}_{-0.00}$ & $-2.56\pm0.01^{+0.01}_{-0.01}$ & $-1.07\pm0.00^{+0.00}_{-0.00}$ & $-2.52\pm0.00^{+0.01}_{-0.01}$   & 0.30 \\
\vspace{2pt}
$V$ & {\bf 6} & - & $0.33\pm0.00^{+0.00}_{-0.00}$ & $-0.82\pm0.00^{+0.00}_{-0.01}$ & $-2.19\pm0.00^{+0.01}_{-0.02}$ & $-1.15\pm0.01^{+0.00}_{-0.01}$ & $-2.52\pm0.01^{+0.01}_{-0.02}$   & 0.24 \\
\vspace{2pt}
$V$ & 7 & $0.90\pm0.03$ & $0.09\pm0.00^{+0.00}_{-0.00}$ & $-1.01\pm0.00^{+0.00}_{-0.01}$ & $-2.31\pm0.00^{+0.00}_{-0.02}$ & $-1.10\pm0.00^{+0.00}_{-0.01}$ & $-2.39\pm0.00^{+0.00}_{-0.01}$   & 0.24 \\
\vspace{2pt}
$V$ & 9 & $0.98\pm0.03$ & $0.04\pm0.00^{+0.00}_{-0.00}$ & $-1.07\pm0.00^{+0.00}_{-0.01}$ & $-2.51\pm0.01^{+0.01}_{-0.02}$ & $-1.11\pm0.00^{+0.00}_{-0.01}$ & $-2.55\pm0.01^{+0.01}_{-0.02}$   & 0.24 \\
\vspace{2pt}
$R$ & {\bf 3} & - & $0.52\pm0.00^{+0.00}_{-0.00}$ & $-0.66\pm0.00^{+0.01}_{-0.00}$ & $-2.15\pm0.00^{+0.04}_{-0.01}$ & $-1.18\pm0.00^{+0.01}_{-0.00}$ & $-2.68\pm0.01^{+0.04}_{-0.01}$  & 0.13 \\
\vspace{2pt}
$R$ & 4 & $1.00\pm0.02$ & $0.53\pm0.01^{+0.00}_{-0.00}$ & $-0.66\pm0.00^{+0.01}_{-0.00}$ & $-2.15\pm0.00^{+0.04}_{-0.01}$ & $-1.20\pm0.01^{+0.01}_{-0.00}$ & $-2.68\pm0.01^{+0.04}_{-0.01}$   & 0.13 \\
\vspace{2pt}
$R$ & 6 & $0.99\pm0.04$ & $0.41\pm0.00^{+0.00}_{-0.00}$ & $-0.76\pm0.00^{+0.01}_{-0.00}$ & $-2.14\pm0.00^{+0.04}_{-0.01}$ & $-1.17\pm0.00^{+0.01}_{-0.00}$ & $-2.55\pm0.00^{+0.04}_{-0.01}$   & 0.13 \\
\vspace{2pt}
$R$ & 7 & $0.90\pm0.04$ & $0.14\pm0.00^{+0.00}_{-0.00}$ & $-0.96\pm0.01^{+0.01}_{-0.00}$ & $-2.23\pm0.00^{+0.04}_{-0.01}$ & $-1.10\pm0.01^{+0.01}_{-0.00}$ & $-2.38\pm0.00^{+0.04}_{-0.01}$   & 0.13 \\
\vspace{2pt}
$R$ & 9 & $0.95\pm0.01$ & $0.10\pm0.00^{+0.00}_{-0.00}$ & $-1.00\pm0.00^{+0.01}_{-0.00}$ & $-2.43\pm0.01^{+0.05}_{-0.01}$ & $-1.10\pm0.01^{+0.02}_{-0.00}$ & $-2.53\pm0.01^{+0.05}_{-0.01}$  & 0.13 \\
\vspace{2pt}
$I$ & 9 & - & $0.01\pm0.00^{+0.00}_{-0.00}$ & $-1.01\pm0.01^{+0.01}_{-0.06}$ & $-2.29\pm0.01^{+0.07}_{-0.05}$ & $-1.02\pm0.01^{+0.01}_{-0.08}$ & $-2.30\pm0.01^{+0.05}_{-0.04}$   & 0.30 \\
\vspace{2pt}
$J$ & 1 & - & $0.34\pm0.00^{+0.02}_{-0.01}$ &$-0.64\pm0.01^{+0.01}_{-0.01}$ &$-2.17\pm0.02^{+0.06}_{-0.03}$ &$-0.99\pm0.01^{+0.02}_{-0.01}$ &$-2.51\pm0.02^{+0.04}_{-0.02}$   & 0.29 \\
\vspace{2pt}
$J\#5133$ & 8 & - & $-0.01\pm0.01^{+0.00}_{-0.00}$ & $-0.84\pm0.01^{+0.04}_{-0.06} $& $-1.78\pm0.01^{+0.06}_{-0.04}$ & $-0.84\pm0.01^{+0.04}_{-0.06}$ & $-1.78\pm0.01^{+0.06}_{-0.04}$  & 0.21 \\
\vspace{2pt}
$F160W$ & 7 & - & $-0.28\pm0.01^{+0.02}_{-0.13}$ & $-1.16\pm0.01^{+0.17}_{-0.10}$ & $-2.57\pm0.05^{+0.05}_{-0.05}$  & $-0.87\pm0.01^{+0.19}_{-0.03}$ & $-2.28\pm0.05^{+0.01}_{-0.02}$  & 0.08 \\
\vspace{2pt}
$H\#5209$ & 8 & - & $-0.21\pm0.01^{+0.00}_{-0.00}$ & $-0.76\pm0.01^{+0.02}_{-0.04} $& $-1.94\pm0.05^{+0.05}_{-0.05}$ & $-0.56\pm0.01^{+0.02}_{-0.04}$ & $-1.73\pm0.06^{+0.05}_{-0.05}$  & 0.21 \\
\vspace{2pt}
$Ks$ & 2 & - & $-0.39\pm0.00^{+0.02}_{-0.00}$ & $-1.00\pm0.01^{+0.01}_{-0.01}$ & $-2.46\pm0.00^{+0.03}_{-0.01}$ & $-0.61\pm0.01^{+0.01}_{-0.04}$ & $-2.06\pm0.00^{+0.03}_{-0.04}$  & 0.21 \\
\vspace{2pt}
$K'\#5337$ & 8 & - & $-0.41\pm0.01^{+0.03}_{-0.03}$ & $-1.01\pm0.01^{+0.01}_{-0.05} $& $-2.28\pm0.04^{+0.08}_{-0.01}$ & $-0.60\pm0.02^{+0.04}_{-0.07}$ & $-1.87\pm0.04^{+0.02}_{-0.06}$  & 0.21 \\

\hline
\end{tabular} 
\label{tab:photom}
\end{center} 
\end{table*}

We report in this section the flux ratios between the lensed images
deduced from the multi-epoch deconvolution when possible and from
single epoch data otherwise (cf Sect.~\ref{subsec:method}). 

Multi-epoch deconvolution consists in deconvolving simultaneously a
set of frames obtained in a filter at several epochs. We report the
results obtained for $B-$, $V-$ and $R-$band data{\footnote{Although $J$ ($K$) band data exist at two epochs,
    the filter response curves are significantly different. Flux variations are
    thus less reliable and not reported.}} at 3, 3 and
5 epochs. For all epochs and all bands, except epoch 6 in the $V$
band, all individual frames have been used. For epoch 6 in the $V$ band, we only 
included the 5 best-seeing frames. For each filter, we rebinned the FORS1
frames to the sampling of the FORS2 frames. Then, we applied frame to
frame zero-point correction calculated based on photometric
measurements of common stars in each field. Then, we ran MCS on the
registered data and derived the magnitude differences and flux time
variations. Due to a small apparent systematic bias ($\sim$0.01 mag)
of the MCS photometry, we corrected our measurements based on the
comparison of the aperture photometry and MCS photometry of 4 field
stars. Final results are presented in Table~\ref{tab:photom} (11 first
rows).

The reduction procedure of single epoch data has already been described in
Sect.~\ref{subsec:absphot}. Only CFHT-IR data have not yet been
discussed. For those, the complete set of individual frames have been
simultaneously deconvolved with a final FWHM of 0.21\arcsec. 

The full results are reported in Table~\ref{tab:photom}. The
photometric (random) errors were calculated based on the dispersion of
the measurements obtained for the different frames. Following this
method, the Poisson noise and the error coming from 
the imperfection of the kernel used for deconvolution are
included in the error bars. This is
confirmed by the fact that estimated errors are larger or equal to
photon noise. The systematic error was estimated from the results obtained for
$\lambda_{\rm low}$ and $\lambda_{\rm high}$ (cf Sect.~\ref{subsec:method}). We observe,
between these two limits, typical variations of the flux ratio by 5 up
to 15\% from the $V$ to the $K$ bands while it is $\leq$ 2\% in the
$B$ band for which the contrast between the ring and the lensed QSO
images is maximum. At a given wavelength, these systematic
errors are reduced when the S/N in the Einstein ring increases.

Finally, note that whenever possible, we have compared the flux ratios
deduced during the multi-epoch process to single epoch deconvolution
results. Both measurements are basically identical{\footnote{Optical single epoch
    magnitude difference can be inferred for epoch 9 in
    Table~{\ref{tab:absphot}}.}} but multi-epoch results are less
sensitive to systematic errors because of the increased S/N in the
ring that improves the separation between point-like and extended flux.


\subsubsection{Caveat}
\label{sec:caveat}

Because of the presence of the Einstein ring superimposed over the
quasar images, one may ask whether the measured point-like fluxes are
associated with the quasar only (i.e. compact emission) or whether they
are significantly mixed with emission originating from a more spatially
extended (partly unresolved) region of the host galaxy. In the
following we examine how the core of the host galaxy may modify the
observed fluxes of images A, B and C.

The observed flux ratios between images A, B, C are contaminated in two
ways by the ring.

The first one is a purely instrumental effect. The {\it resolved} ring
plays the role of a spatially variable background, whose intensity is
increasing below the unresolved PSF because this region corresponds to
the innermost parts of the host. However the deconvolution process
tends to produce a "flat" background below the PSF whose intensity
closely matches the value in its close vicinity. Since the surface
brightness of the {\it resolved} background ring is the same around
each point-like image, the flux of the faint images (i.e. weakly
amplified) is more contaminated by the ring than the flux of the bright
ones. Consequently, the flux {\it ratios} between brighter and
weaker lensed images are {\it underestimated} with respect to the
theoretical values. This ring contamination increases with the ring
brightness (i.e. from the $B$ band to the $K$ band) and we call this the
{\it background effect}. 

    \begin{figure}[tb] 
   \centering 
\resizebox{0.9\linewidth}{!}{\includegraphics{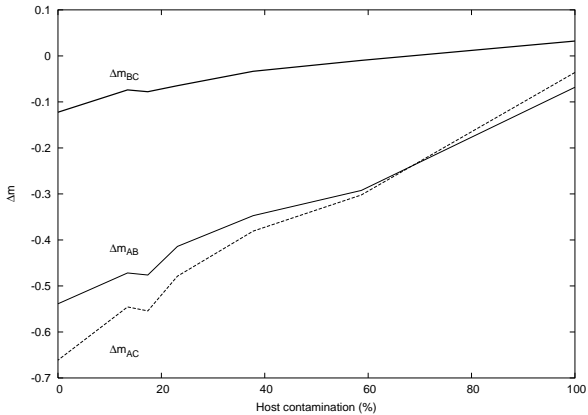}} 
\caption{Simulated point-like magnitude difference $\Delta
  m_{AB}$(thin solid line), $\Delta m_{AC}$ (dotted line) and $\Delta
  m_{BC}$ (bold solid line) as a function of the average fraction of the flux of the host galaxy included in the total point-like flux after deconvolution (host contamination).}
\label{fig:contrast} 
    \end{figure}

The second effect is related to the specific geometry of J1131 in the
source plane. Indeed, according to the lens models, the QSO core is
located at only 0.016\arcsec (115 pc in the source plane; Paper I)
from a macro caustic. Hence, the dusty torus and the most central part
of the host galaxy (that contributes significantly to the NIR flux) cover a
region in the source plane with a strong magnification
gradient. Therefore, the amplifications of the lensed images will
depend on the considered source radius even at very small, unresolved
scales. Moreover, since the local magnification around image A is
larger than around images B \& C, the {\it unresolved} source radius will
be smaller for image A and the amplification {\it ratios} will also be
affected. We dub this effect the {\it differential amplification effect}.

Although the latter effect depends on the radial distribution of the
source light, which is itself a function of wavelength, we made
simulations to quantify its trend. We first chose a circularly
symmetric exponential light profile with a half-flux radius $r_e=1.9$
kpc (= 0.25\arcsec) to represent a typical quasar host galaxy
(S\'anchez et al. \cite{SAN04}). We lensed that extended source, 
    sampled on a grid of 0.00025\arcsec/pixel, with a Singular Isothermal
Ellipsoid + shear model fitting our data (SIE+$\gamma$;
Sect.~\ref{sec:mod}) using the {\texttt {gravlens}} software v1.06
(Keeton, \cite{KEE01}). The image plane, sampled on a 5 mas
    grid, was convolved with a Moffat profile (FWHM=0.7\arcsec) and
rebinned to a scale of 0.16\arcsec/pixel to mimic the observations.
This synthetic frame was then deconvolved using the MCS technique to a
final seeing of FWHM = 0.32\arcsec. We repeated this operation for
several source radii in the range $ 0.005$\arcsec$ < r_{\rm src} <
0.1$\arcsec and we computed the flux ratios I$_{\rm A}$/I$_{\rm B}$
and I$_{\rm A}$/I$_{\rm C}$. The main result is that the source
remains unresolved in the A, B and C images for $r_{\rm src} <
0.015$\arcsec. The associated flux ratios are then nearly identical to
the point-like source ones ($\Delta m \sim 0.05$ mag). For larger
source radii, the images become resolved and these flux ratios {\it
  decrease} and converge towards 1.

Finally, we simulated a more realistic source model with two
    components: an extended exponential profile (truncated to $r_{\rm
      host} = 0.1$\arcsec) lensed into a resolved background ring and
    a superimposed disk-like source ($r_{\rm QSO} = 0.01$\arcsec)
    lensed into unresolved images. This enabled us to look for the
    {\it joined influence} of the {\it background effect} and the {\it
      differential amplification effect} on the observed magnitude
    differences between the unresolved images. We quantify the effect as a
    function of the host contamination (defined as the fraction of the
    total point-like flux coming from the host galaxy). The results
are displayed in Fig.~\ref{fig:contrast}.

Because the QSO Spectral Energy Distribution (SED) is expected to be dimmer than the host SED in the
NIR range, we may expect a stronger ring contamination in that
range. An accurate estimate would require us to know the true SEDs,
but we can already infer an approximate contamination from the existing
data. Indeed, in the $B$ band, the ring image is faint and we estimate
the host contribution to be at maximum 5\% of the total point-like
flux. Using a mean color index $B-K=2.5$ for the QSO and $B-K=5$ for
the host galaxy, we calculate that in the $K$ band up to 35\% of the
total point-like flux is coming from the host. Reporting these values
in Fig.~\ref{fig:contrast}, and comparing them with the case of 0\% contamination, we derive that the observed flux ratios should not
deviate by more than 0.05 mag from the predictions for a point-like
source in the $B$ band. In the $K$ band, $\Delta m_{\rm AB}$ and
$\Delta m_{\rm AC}$ are reduced by several tenths of magnitude. On the
contrary, $\Delta m_{\rm BC}$ does not vary significantly.  


\section{Results}
\label{sec:results}

\subsection{Ring morphology}
\label{subsec:ring}

The multi-wavelength images of J1131 confirm that the contrast between
the QSO and its host decreases with increasing wavelength, as expected
for the host galaxy of a quasar located at $z\sim0.66$. Indeed, the
SED of QSOs is minimum at $\sim$ 1-2
$\mu$m (rest-frame) while in this range the dominant emission is
coming from the host (e.g. Dunlop et al. \cite{DUN93}) and marginally
from the dust torus (hot dust emission increases up to $\sim$ 2
$\mu$m; e.g. Elvis et al. \cite{ELV94}, Nenkova et
al. \cite{NEN02}).

\begin{figure*}[tb!] 
\centering
\resizebox{0.7\linewidth}{!}{\includegraphics{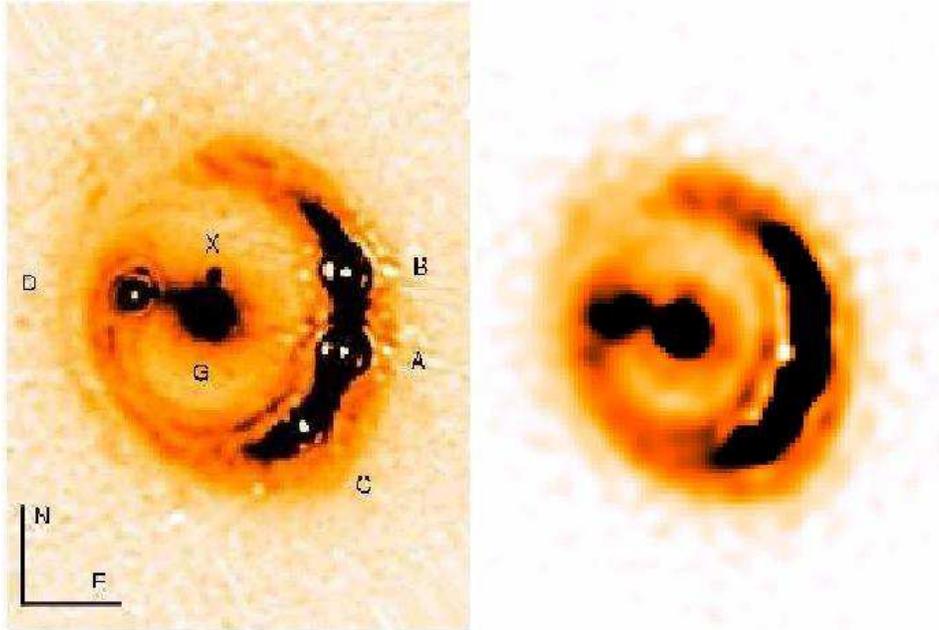}}
\caption{{\it Left} :  Deconvolved HST NICMOS $F160W$ image mainly showing the ring. FWHM=0.08\arcsec. {\it Right} : Deconvolved image in the $Ks$ filter obtained with ISAAC. FWHM=0.15\arcsec}
\label{fig:MCSHK} 
    \end{figure*}

While some structures in the ring are barely detected on the ground
based $Ks$ deconvolved frames (Fig.~\ref{fig:MCSHK}, right), the HST NIC2 deconvolved image
(Fig.~\ref{fig:MCSHK}, left) spectacularly unveils many details in the
ring. Additionally, a putative companion object ({\it X}) near the
lensing galaxy is more clearly detected. Its nature is investigated in
Paper II based on ACS and NICMOS data and it is shown that the
expected influence of X on the lens modeling can only be very weak.

\subsection{Lens modeling}
\label{sec:mod}

In this section we construct a fiducial lens model aimed at discussing
the observed flux ratios and estimating the time delays for J1131. This
adopted lens model is a Singular Isothermal Ellipsoid (SIE) +
external shear ($\gamma$), which is generically used in the study of
the flux ratio anomalies (Dobler \& Keeton~\cite{DOB05} and references
therein). It has also the advantage to be fully constrained by the
astrometry of J1131 (i.e. the positions of A, B, C, D and G).
Additionally, it takes into account the main characteristics of the
deflector, namely the lens galaxy ellipticity and the external shear
associated with the lens environment. We used the {\texttt {gravlens}}
modeling package v~1.06 (Keeton, \cite{KEE01}) constrained by the HST
relative astrometry of the point-like images and of the lensing galaxy
(Sect.~\ref{subsec:relastr}). Since systematic errors affecting the
NICMOS data (error on the plate scale, on the distorsion, on the
orientation and due to PSF undersampling) do not influence similarly
all lensed images, we have included them in the position error bars.
The model yields the following results : $\theta_E =$ 1.843\arcsec, $e=$
0.162, $\theta_e =$ -57.92$\degr$, $\gamma=$ 0.114 and
$\theta_{\gamma}=$ -83.01$\degr$; where $\theta_E$ is the Einstein angular
radius, $e$ is the ellipticity and $\theta_e$ is its PA, $\gamma$ is
the tidal shear and $\theta_{\gamma}$ (counted E of N) points towards
the mass producing the shear. We should note that the shear does not point
  towards any obvious massive perturber in the field. On the other hand $e$ and
  $\theta_e$ agree well with the light profile ellipticity and P.A. of 
  the lens galaxy (see Paper II). The reduced $\chi^2$ (for 1 d.o.f.)
associated with this model amounts to 136 which is statistically
unacceptable. The disagreement between the model predictions and the
observations is mainly caused by the lensing galaxy position ($\chi^2_{\rm
  gal}=$ 90) expected $0.03\arcsec$ South from its observed position
(Table~\ref{tab:summary}). 
  
Assuming that we can ignore X in the modeling, we opted for the
SIE+$\gamma$ model as our fiducial lens model{\footnote{We also
    emphasize that SIE+$\gamma$ left us with only 1 d.o.f. More complex 
    models thus need additional constraints (see Paper II).}}. This model leads to
predicted time delays between C (leading) and the other multiple
images : $\tau_{\rm   CB} = 0.3$ d, $\tau_{\rm CA} = 1.3$ d and
$\tau_{\rm CD} = 126.9$ d. The predicted flux ratios are reported in
Table~\ref{tab:summary}. We have also fitted an ensemble of non
parametric asymmetric models (searched by a Monte-Carlo method) with
the {\texttt{PixeLens}} publicly available software (Saha \&
Williams~\cite{SAH04}). This enabled us to derive a distribution of
$\tau_{\rm   CD}$ (each  $\tau_{\rm   CD}$ being associated with a
different model) in the range 80-400 days while $\tau_{\rm CA}$ may
not exceed 5 days. 

\subsection{Flux variability}
 \label{subsec:BVR}

\subsubsection{Temporal flux ratio variations}
\label{subsec:temp}

    \begin{figure}[] 
   \centering 
\resizebox{\linewidth}{!}{\includegraphics[bb = 15 45 405 310, clip]{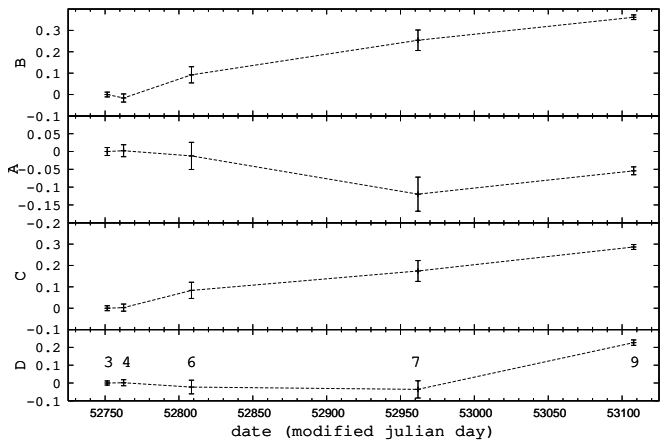}} 
\caption{Lightcurves of images B, A, C and D in the $R$ band for epochs
  3, 4, 6, 7, 9. The x axis refers to the observing date in modified
  julian day (i.e. julian day-2400000). The y axis refers to the
  magnitude difference with respect to epoch 3 (i.e. $m_i - m_3$). The
  dotted line joining the points is plotted for the legibility of the graph. }
\label{fig:lightR} 
    \end{figure} 

Flux variations of the lensed images of J1131 have been obtained in
Sect.~\ref{subsec:relphot}. Figure~{\ref{fig:lightR}} represents the
 corresponding lightcurves in the $R$ band{\footnote{One might suspect
     that the difference of response between the Special $R$
filter (epochs 3, 4, 6, 7) and the Bessel $R$ filter (epoch 9) biases the observed
lightcurve. But the color correction only amounts to 0.02 mag for the spectrum of J1131 and should not significantly modify
our results.}}. The comparison of these
 ones with the lighcurves observed in the $V$ and $B$ bands (i.e. epochs 6-7-9 in
$V$ and epochs 7 and 9 in $B$) indicates an agreement
between the lightcurves within 0.05 mag and often
around 0.02 mag. Lightcurves are also nearly insensitive to the choice
of the regularization parameter (except in the $R$ band where the
maximum shift of the D lightcurve can reach 0.05 mag).

Figure~{\ref{fig:lightR}} reveals a dimming of B and C by about 0.3 mag
 in an eleven month period. A dimming of D is observed with a possible 
 delay of about 5 months. During this period, the flux of A changed by only
 $\sim$ 0.1 mag. 

Because the time-delay between A-B-C is of the order of a few days,
intrinsic flux variations of the source would produce nearly
simultaneous flux variations between these images. Therefore, the different
behaviour of image A with respect to B and C suggests that a
microlensing event occurs in image A (i.e. microlensing partially
compensates the intrinsic flux variations; scenario
S1). Alternatively, B and C might be dimmed by two independent
microlensing events producing nearly simultaneously the same
lightcurve for both images (scenario S2). For both S1 and S2,
the observed microlensing time-scale is compatible with the crossing
time of a microcaustic by the source (e.g. Treyer \&
Wambsganss~\cite{TRE04}). Finally, we note that the dimming of D five
months (i.e. $\sim \tau_{CD}$) after B and C supports S1, but
microlensing in D cannot either be excluded. 

\subsubsection{Chromatic variations}
\label{subsec:flux}

Three phenomena can {\it a priori} introduce chromatic effects in the
observed flux ratios: {\it (i)} differential extinction produced by
dust in the lensing galaxy and which is expected to mainly affect the blue bands;
{\it (ii)} microlensing, which is stronger for smaller source sizes
(e.g.Wambsganss \& Paczy\'nski \cite{WAM91}, Wyithe et
al. \cite{WYI00}) and therefore for the shorter wavelengths emitted in 
the most central parts of the accretion disk (in the NIR, the microlensing
signal would also be swamped by the constant flux contribution from
the host); {\it (iii)} systematic errors due to the {\it differential
amplification} and to the {\it background effect} which decrease by
several tenths the modulus of the magnitude difference between two
images from B to K (Sect.~\ref{sec:caveat} and Fig.~\ref{fig:contrast}).

    \begin{figure}[] 
    \centering 
\resizebox{\linewidth}{!}{\includegraphics[bb=15 40 405 310]{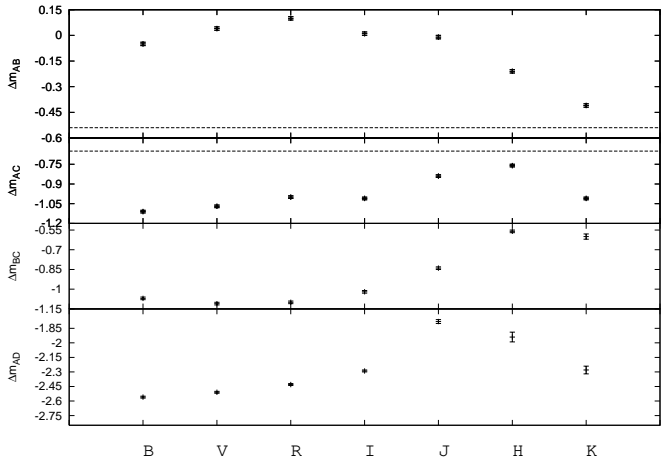}}
\caption{Magnitude differences between A (or B) and the other
  point-like images in $B$, $V$, $R$, $I$ as observed at epoch 9 and in $J$, $H$,
  $K$ at epoch 8 (cf Table~\ref{tab:photom}). The dotted line, when
  visible, represents the flux ratio predicted by the SIE+$\gamma$ model,
  namely $\Delta m_{\rm AB}$ = -0.54, $\Delta m_{\rm AC}$ = -0.65,
  $\Delta m_{\rm BC}$ = -0.11 and $\Delta m_{\rm AD}$ =
  -3.23. 1$\sigma$ error bars are displayed.  }
\label{fig:DeltaAi} 
    \end{figure}

Figure~\ref{fig:DeltaAi} shows the magnitude differences between
    the point-like images as seen through different filters from the
    $B$ to the $K$ band, at epochs 8 (NIR
    range) and 9 (optical range). Chromatic variations are
    low (i.e. $\leq$ 0.15 mag) in the optical while variations up to
    0.5 mag are observed within the NIR range. This first statement does
    not argue in favor of strong differential extinction between
    lensed images since this one should be stronger in the optical
    range (where effect {\it iii} is negligible) than in the NIR.
    Differential reddening can thus be safely neglected in the
    following.
    
    We now interpret the chromaticity of $\Delta m_{\rm AB}$, $\Delta
    m_{\rm BC}$ and $\Delta m_{\rm AC}$ (magnitude differences
    involving D are not discussed because of the large unknown time
    delay for this image) between optical and NIR wavelengths. Note
    that in the following, we discuss the chromatic variations for
    optical and NIR range separately when microlensing might have
    produced flux ratio variations between epochs 8 -NIR- and 9
    -optical-:
\begin{itemize}
\item $\Delta m_{\rm AB}$: Figure~\ref{fig:DeltaAi} shows that
  $\Delta m_{\rm AB} \sim 0$ mag in the optical and decreases down to
  -0.45 mag in the $K$ band. Both microlensing scenarii described in
Sect.~\ref{subsec:temp} can reproduce the fact that $|\Delta m_{\rm
  AB}|$ is smaller in the $J$ band than in the $K$ band: either A is
affected (S1) and is then {\it de-amplified} or B is affected (S2) and
is then {\it   amplified}. Note that we do not expect chromatic
variation caused by phenomenon {\it iii} since $\Delta m_{\rm AB} \sim 0$ mag
in the $J$ band. Whatever the adopted scenario, the value of
$I_{\rm B}/I_{\rm A}$ observed in the $K$ band is the least affected by 
microlensing and thus constitutes an upper limit on the true flux 
ratio $I_{\rm B}/I_{\rm A}$ (Table~\ref{tab:summary}). 
\item $\Delta m_{\rm BC}$: $|\Delta m_{\rm BC}| \sim 1$ mag in the
  optical range and decreases to $|\Delta m_{\rm BC}| \sim 0.55$ mag
  in the $H$ and $K$ bands. Under S1, this fading of I$_{\rm
        B}/$I$_{\rm C}$ from the $B$ to the $K$ band can be explained by effect {\it iii} (since $|\Delta m_{\rm BC}|$ differs
      significantly from 0 and B and C are not microlensed).
Note that since $\tau_{\rm CB}$ is very short and
  since the ring contamination is negligible in the $B$ band, $\Delta
  m_{\rm BC}$ in the $B$ band should correctly represent the
  macro-lens flux ratio under S1. However, this value ($\Delta m_{\rm
    BC}$ = -1.05 mag) is abnormally low compared to the value
  predicted by the model ($\Delta m_{\rm BC}^{model}$ = -0.11 mag).
  Under scenario S2, several combinations of microlensing for B and C
  and effect {\it iii} are possible.
\item  $\Delta m_{\rm AC}$ : $|\Delta m_{\rm AC}| \sim 1.05$ mag in
  the optical range. It amounts to 0.84 mag in the $J$ band, 0.76 mag in
  the $H$ band and rises up to 1.01 mag in the $K$ band. First
  under S1, we can explain the observed decrease of $|\Delta m_{\rm
    AC}|$ between the $J$ and $H$ bands by the dominant ring
  contamination, while the subsequent increase of $|\Delta m_{\rm
    AC}|$ in the $K$ band would be due to the drop of the microlensing
  {\it de-amplification} of image A. Under S2, the same
  interpretation remains valid except that the increase of $|\Delta
  m_{\rm AC}|$ in the $K$ band implies that image C is more {\it
    amplified} by microlensing at bluer wavelengths. In the latter
  case, its value in the $K$ band should be a lower limit of the
  macro-lens flux ratio. This reveals an abnormally large value of
  $|\Delta m_{\rm AC}|$ (compared to the fiducial model
  prediction) under scenario S2.

\end{itemize}

In summary, our data first suggest the absence of strong differential
extinction between the lensed images of J1131. Second, the observed chromatic
variations of the flux ratios can qualitatively be explained by the
combination of effects {\it ii} and {\it iii}. More interestingly,
this analysis also reveals that either A is {\it de-amplified} by a
microlensing event (S1) or B and C are both {\it amplified} by two
independent microlensing events (S2). For both microlensing scenarii,
we have explained in which filter the flux ratio measurement should be
the closest to the values predicted by the ``macro model''. As one can
see in Table~{\ref{tab:summary}}, these ``best'' estimates of the flux
ratios do not reconcile with the macro model predictions. We investigate
how {\it anomalous} are the flux ratios in the next section.

\subsubsection{Anomalous flux ratios}
\label{sub:summarydiscuss}

\begin{table*}[tb]
\caption{Summary of the observed relative astrometry and photometry
  related to J1131 and comparison with model predictions. The observed relative
  positions with respect to image A ({\it{obs}}), obtained from NICMOS
  images, are those given in
  Table~\ref{tab:astro}. The systematic errors on the relative
  positions have been quadratically added to the random errors
  affecting the measurements. Since the position of A is not fixed to
  the observed one but known with the uncertainty reported in
  Table~{\ref{tab:astro}}, the modeled position of A is not
  (0,0). Flux ratios $I_i/I_j$ in the $K$ band are reported when
  microlensing (scenario S1 or S2) affects image $i$ or $j$ and in the
  $B$ band otherwise. Since microlensing is minimum in the $K$ band,
  the flux ratio reported is a lower/upper limit on the ``true flux
  ratio''. Flux ratios including image D are placed into brackets
  because they are not corrected for the unknown time
  delay. Additionally, a microlensing event may also occur for that image. }

\begin{center}
\begin{tabular}{c|rc|cc|rc|rc}
\hline 

ID (Parity) & \multicolumn{2}{c}{$\Delta \alpha \cos \delta('')$} &
\multicolumn{2}{c}{$\Delta \delta ('')$} & \multicolumn{2}{|c}{I$_{\rm
    B}/$I$_j$} & \multicolumn{2}{c}{I$_{\rm A}/$I$_j$} \\
\hline 
& obs & mod$-$obs & obs & mod$-$obs & S1 & mod & S2 & mod\\
\hline
A ($-$) & $0.000\pm0.003$  & -0.002  & $0.000\pm0.003$  & -0.002  &
$\leq$ 0.68 	  & 0.61 & 1.00              & 1.00      	\\
B ($+$) & $0.032\pm0.003$  & -0.002  & $1.184\pm0.003$  & 0.003   &
1.00 		  & 1.00 & $\geq$ 1.46       & 1.65      	\\
C ($+$) & $-0.590\pm0.003$ & 0.002  & $-1.117\pm0.003$ & 0.003    &
$2.67\pm0.01$	  & 1.11 &  $\geq$ 2.53      & 1.83       \\ 
D ($-$) & $-3.115\pm0.003$ & 0.006  & $0.875\pm0.004$  & 0.019    &
[$10.02\pm0.01$]	& 11.95	 & [$10.58\pm0.06$]  & 19.72      \\
G       & $-2.027\pm0.003$ & -0.010  & $0.607\pm0.004$  & -0.029  & -
& - 	   & -               & -          \\
X       & $-1.936\pm0.011$ & -       & $1.100\pm0.011$   & -       & -
& - 	   & -               & -         \\

\hline
\end{tabular} 
\label{tab:summary}
\end{center} 
\end{table*}

Anomalous flux ratios in lensed systems have caught little attention until the
work of Mao \& Schneider (\cite{MAO98}) who first realised that
substructures could significantly modify the flux ratios in
B1422+231. Various authors tried to compare the amount of Dark Matter
Halo substructures predicted by $\Lambda$CDM hydrodynamical
simulations to the amount of substructures necessary in lensing models to reproduce observed flux
ratios (e.g. Metcalf \& Madau~\cite{MET01},
Chiba~\cite{CHI02}, Dalal \& Kochanek~\cite{DAL02}, Metcalf \&
Zhao~\cite{MET02}, Mao et al.~\cite{MAO04}, Brada{\v c} et
al.~\cite{BRA02},~\cite{BRA04}, Kochanek \&
Dalal~\cite{KOC04}, Metcalf~\cite{MET05}). However, it is still unclear whether the amount of 
substructures necessary in lensing models agrees or not with
the hydrodynamical simulation predictions (see e.g. KGP05 or Rozo et al.~\cite{ROZ05}
for a recent extensive summary of the situation). 

Indeed, from a lensing perspective, flux
ratios are not only sensitive to substructures but also to m=3 and
m=4 multipoles of the lens model (Evans \& Witt~\cite{EVA03}, M\"oller
et al.~\cite{MOL03}, Quadri et al.~\cite{QUA03}, Kawano et
al.~\cite{KAW04}), to the lens environment (Oguri~\cite{OGU05}), to the
source size (Dobler \& Keeton~\cite{DOB05}, Inoue \&
Chiba~\cite{INO04}, Chiba et al.~\cite{CHI05}) and to isolated dark
matter clumps on the same line of sight (Chen et al.~\cite{CHE03},
Metcalf~\cite{MET05}, ~\cite{MET05b}, Wambsganss et
al.~\cite{WAM04}). From a hydrodynamical simulation perspective, the expected amount of dark
matter within the Einstein radius of a lens is not well known. Indeed,
both the abundance of halo substructures (e.g. Zentner \&
Bullock~\cite{ZEN02}, Sigurdson \& Kamionkowski~\cite{SIG04}) and
their spatial distribution (e.g. Chen et al.~\cite{CHE03}, Zentner \&
Bullock~\cite{ZEN03}, De Lucia et al.~\cite{LUC04}, Oguri \&
Lee~\cite{OGU04}) are uncertain. In particular the amount of
dark matter clumps that survives in the central part of a halo (and so
within a lens Einstein radius) is still debated (e.g. Mao et
al.~\cite{MAO04}, Amara et al.~\cite{AMA04}, Trott \& Melatos
~\cite{TRO05}, Macci\`o et al.~\cite{MAC05}).
 
In any case, the first step of any analysis is to obtain secure flux
ratios and to identify anomalies independently of the choice of a
macro-lens model. For this purpose, the most reliable method is based on
the validity/violation of the so called ``magnification sum rules'',
originally introduced by Schneider \& Weiss (\cite{SCH92a}). The first
rule predicts that a pair of images created  when the source crosses a
fold caustic will have the same brightness. The second one is valid
when an {\it unresolved} source lies near a cusp caustic and predicts
that the quantity $R_{\rm cusp}$ (Mao \& Schneider~\cite{MAO98}) : 

\begin{equation}
R_{\rm cusp} = {\frac{{|{\mu_{\rm A}+\mu_{\rm B}+\mu_{\rm C}}|}}{{|\mu_{\rm A}|}+{|\mu_{\rm B}|}+{|\mu_{\rm C}|}}} = {\frac{{|{1-I_{\rm B}/I_{\rm A}-I_{\rm C}/I_{\rm A}}|}}{1+I_{\rm B}/I_{\rm A}+I_{\rm C}/I_{\rm A}}}, 
\label{equ:rcusp}
\end {equation}

{\noindent} converges towards 0 as the source gets closer to a cusp
singularity and all the three images are merging.

The latter relation appears to be more robust than the relation
binding ``fold images'' (KGP03, KGP05). Cusp configuration systems are 
thus particularly important for the study of flux ratio anomalies (KGP03,
Brada{\v c}~\cite{BRA04}). Among those, only two systems possess radio flux
ratios (B1422+231 and B2045+265{\footnote{We do not consider the
    hybrid fold/cusp case of B0742+472.}}). The two others (i.e. RXJ
0911+0551 and J1131) could only be observed in the optical range and 
may thus be contaminated by dust extinction and
microlensing. Hereafter, we further discuss the case of J1131 for
which $R_{\rm cusp}\sim$ 0.35 at the epoch of the discovery (e.g. KGP03). 

Although microlensing precludes the determination of the true flux
ratios, we have shown in Sect.~\ref{subsec:flux} that reliable
upper/lower limits on these ones could be derived
(Table~{\ref{tab:summary}}). Using the flux ratios of
Table~{\ref{tab:summary}}, we find that $R_{\rm cusp} \geq$ 0.036
whatever the microlensing scenario. Although $R_{\rm cusp}$ might
naturally deviate from 0 (KGP03), the observed value is fully
consistent with no violation of the cusp relation. This does not
automatically mean that effects of substructures are rejected. First,
we only have a lower limit on $R_{\rm cusp}$ and second, ``small''
$R_{\rm cusp}$ values can be observed while flux ratios cannot be
reproduced by realistic lens models (cf B1422+231; KGP03, Brada{\v c}
et al.~\cite{BRA02}). 

As already noticed in Sect.~\ref{subsec:flux}, ${I_{\rm B}}/{I_{\rm
    C}}$ (${I_{\rm A}}/{I_{\rm C}}$) does not agree with our fiducial
model predictions under S1 (S2). Remembering that substructures can only {\it magnify} a
minimum of the arrival time surface and {\it preferentially demagnify} 
saddle points (e.g. Schechter \& Wambsganss~\cite{SCH02}), we now look 
if substructures are able to explain that flux ratio. Under scenario S1, ${I_{\rm B}}/{I_{\rm C}}=$
2.67 while the predicted value is 1.11 (Table ~\ref{tab:summary}). Both B and C are minima of the
arrival time surface. Consequently, only a magnification of B by 0.95
mag is allowed by the substructure scenario (since C being a minimum
cannot be de-amplified by milli-lensing) to explain  ${I_{\rm
    B}}/{I_{\rm C}}$. However, the value of ${I_{\rm A}}/{I_{\rm C}}$ (${I_{\rm A}}/{I_{\rm C}} \geq$ 
3.92) derived under S1 would then imply that A should be {\it
  magnified} by a factor $\geq 2.15$ by substructures to match the
predicted value of  ${I_{\rm A}}/{I_{\rm C}}=1.83$ and this
is very unlikely because i- {\it two} substructures are
    required, ii- the saddle-point image A must be {\it amplified}. \\
Under S2, ${I_{\rm A}}/{I_{\rm C}}\geq$ 2.53 while the macro model predicts
1.83. Since image C is located at a minimum of the arrival time
surface and, thus, cannot be demagnified by substructures, the above
flux ratio can only be explained if image A is amplified by
substructures by more than 0.35 mag
(i.e. 2.5$\log($2.53/1.83$)$). This would also reproduce the observed
value of ${I_{\rm A}}/{I_{\rm B}}$. Nevertheless, since image A is a
saddle-point, it has a low probability to be magnified by a
substructure.

Consequently, under S1, a single substructure in front of a lensed
image cannot explain the observed value of $I_{\rm B}/I_{\rm
  C}$. Under S2, a milli-lensing magnification of A might explain
the observed flux ratios. However, magnification of a saddle point 
by a substructure is statistically quite unlikely. 

\section{Summary and conclusions}
\label{sec:conclusions}

We have presented direct optical and NIR imaging of the quadruply
imaged quasar RXS J113155.4-123155 obtained with ground-based
telescopes and with HST at various epochs. The HST NICMOS images
unveil many details in the Einstein ring of this system as well as a
putative companion object (X) lying at $\sim$1\arcsec~from the lensing
galaxy. Accurate positions have been derived for the 4 lensed images,
for X and for the lensing galaxy based on the HST NIC2 data and on the
ground-based VLT observations. Both sets of measurements do agree within 6
mas. A SIE+$\gamma$ model has been fitted to the observed image
configuration. Nevertheless, the positions of image D and of the lens galaxy
do not match satisfactorily the observed ones. Additionally, we
have used the SIE+$\gamma$ model to derive the expected point-like
flux ratios when taking into account the extended nature of the
source. 

The images have been deconvolved using the MCS method known to
preserve photometry. Hence, we derived for the first time relative and
absolute photometry of J1131 from the optical to the NIR
range. Since the photometric measurements of the point-like images
in J1131 are complicated by the superimposed Einstein ring,
we have thoroughly discussed the potential sources of photometric
errors. We have shown that the MCS method allows to reach a
photometric accuracy close to the photon noise limit in the $B$ and
$V$ filters if a good quality PSF can be constructed. On the other hand,
small systematic errors affect the flux ratios in the $R$ band due to
the Einstein ring.

The small chromatic variations of the flux ratios between the
    $B$ and $R$ bands suggest a low level of differential extinction between the
    lensed images of J1131. 
Based on sparse photometric measurements obtained between November
2002 and April 2004, we have shown the likely presence of microlensing
for image A together with intrinsic quasar flux variability up to 0.3
mag. Microlensing of A is also supported by our Chandra observations 
    of J1131 in the X-rays (see Paper III, Claeskens et al., in
    preparation; Blackburne et al.~\cite{BLA05}). Additionally, based on the chromatic variations of the flux
ratios, we have shown that A should likely be de-amplified (scenario
S1), in agreement with the
fact that it is a saddle-point of the arrival-time surface{\footnote{During the 
     referee process, Rozo et al. (\cite{ROZ05}) demonstrated that
     small scale structures, if they can be considered as linear
     perturbations, might not preferentially demagnify saddle
     point images. However, large perturbations from point-mass
     microlenses are likely caused by non linear perturbers and the
     results of Rozo et al. (\cite{ROZ05}) may not be valid in such a
     case.}} (Schechter \&
Wambsganss~\cite{SCH02},  Keeton~\cite{KEE03a}). Alternatively, our data cannot rule out that
two independent microlensing amplifications (scenario S2) with roughly the same
amplitude occur for both images B and C. 

We have used the $R_{\rm{cusp}}$ relation (Mao \& Schneider, \cite{MAO98};
Eq.(~\ref{equ:rcusp})) to investigate the evidence for substructures in
the lensing galaxy. Once lower/upper bounds on the flux ratios are
imposed from our knowledge of microlensing occuring in this system, we
find a lower limit on $R_{\rm{cusp}}$ close to zero. This shows
that the small scale structure(s) identified in KGP03 from the
$R_{\rm{cusp}}$ relation, was (were) indeed due to (a) microlens(es).

Nevertheless, the observed flux ratios do not agree with the model
predictions. We argue that milli-lensing effects cannot provide the
best explanation under S1 while under S2, observations suggest
that a massive substructure should {\it magnify} the saddle point
image A. This is however a low probability scenario (Schechter \&
Wambsganss~\cite{SCH02}). An alternative to the milli-lensing
explanation is to introduce in the macro-lens potential higher
order multipole terms (e.g. Evans \& Witt \cite{EVA03}). This is described
in Paper II, where we successfully added an octupole ($m=4$) to the fiducial
model and reproduced both the observed flux ratio $I_{\rm B}/I_{\rm C}$
(under S1) and the observed relative lens position. Therefore, we found  that  the
conjunction of microlensing and of higher order terms in the lens
potential could explain the flux ratios observed in J1131. The
preference of a more complex macro model over the presence of
substructure is opposite to what Kochanek \& Dalal (\cite{KOC04}) have
found for other lenses. Of course the solution of adding an octupole
might not be unique and other models such as those including several
sources of shear (e.g. Morgan et al.~\cite{MOR05}), or with two mass
components (e.g. Dye \& Warren~\cite{DYE05}) should be explored.


 \begin{acknowledgements}
   
   Dominique Sluse acknowledges support from an ESO studentship in
   Santiago and PRODEX (Gravitational lens studies with HST). JFC
   acknowledges support from PRODEX (XMM and HST). Part of the
   research was also performed in the framework of the IUAP P5/36
   project, supported by the OSTC Belgian Federal services. We
   gratefully thank M. Bremer, M. Pierre and J. Willis who allowed us
   to perform extra observations of J1131. We also want to thank the
   referee, Steve Warren, whose comments improved the paper. D.S.
   wants to thank Thierry Forveille and Audrey Delsanti for their
   advice in the reduction of the NIR data, Olivier Hainaut for
   helpful discussions on data reductions (among other things),
   Fr\'ed\'eric Courbin for providing us the MCS code. Our ISAAC
   service astronomer, A. Jaunsen, as well as the FORS service
   astronomers are warmly acknowledged for their excellent work.

 \end{acknowledgements}

 \end{document}